\def\la{\mathrel{\mathpalette\fun <}}
\def\ga{\mathrel{\mathpalette\fun >}}
\def\fun#1#2{\lower3.6pt\vbox{\baselineskip0pt\lineskip.9pt
  \ialign{$\mathsurround=0pt#1\hfil##\hfil$\crcr#2\crcr\sim\crcr}}}
\relax
\documentstyle[12pt]{article}
\advance\textheight by \topskip
\begin{document}

\newcommand{\DD}[2]{\mbox{$\frac{\partial^{#1}}{\partial #2}$}}
\newcommand{\Ddd}[2]{\mbox{$\frac{\partial^{2}}{\partial #1 \partial
#2}$}}
\newcommand{\uv}[2]{\mbox{$u(#1,#2,z,{\cal D})$}}
\newcommand{\Dme}{\mbox{$m_{1}(t,{\cal D}|\chi)$}}
\newcommand{\Dma}[1]{\mbox{$m_{1}(t,{\cal D_{#1}}|\chi_{#1})$}}
\newcommand{\EV}[1]{\mbox{${\bf{\sf E}}\left\{ \left. #1 \right| \,
\phi_0 =
\chi \right\} $}}   \newcommand{\Pes}[1]{\mbox{$\psi_{#1}(\chi) $}}
\newcommand{\Dpi}[1]{\mbox{$\pi_{#1}(\phi) $}}
\newcommand{\tW}[1]{\mbox{$\tilde{W}^{#1}(\phi) $}}
\newcommand{\TW}[1]{\mbox{$\tilde{W}^{#1}(\phi(z)) $}}
\newcommand{\Tw}[1]{\mbox{$\tilde{W}^{#1}(z) $}}

\phantom{.}
\rightline{SU-ITP-93-2}
\rightline{gr-qc/9302009}
\rightline{January 29, 1993}
\vskip 0.5cm
\begin{center}
{\Large\bf TOWARDS THE THEORY OF \\
\vskip 0.5cm
STATIONARY UNIVERSE}\footnote{Talk, presented at the TEXAS/PASCOS
conference, Berkeley, December 1992.}\\
\vskip 1.0cm
{\large  \sc Arthur Mezhlumian}\footnote{On leave from: Landau Institute
for Theoretical Physics, Moscow. \ E-mail: arthur@physics.stanford.edu}
 \vskip 0.5cm
{\it Department of Physics, Stanford University, \\
\vskip 0.3cm
Stanford, CA 94305-4060}\\
\end{center}
\vskip 1.0cm

\begin{center}
ABSTRACT

\end{center}

\begin{quote}
This talk presents some progress achieved in collaboration with A.Linde and
D.Linde$^{\ref{LLMPascos},\ref{MezhLin}}$ towards understanding the true
nature of the
global spatial structure of the Universe as well as the most general
stationary
characteristics of its time-dependent state with eternally growing total
volume.
\end{quote}

\vskip 1.5cm
In our opinion, the simplest and, simultaneously, the most
general  version of  inflationary cosmology  is the chaotic inflation
scenario$^{\ref{b17}}$.   It can be realized in all models were the other
versions$^{\ref{b16},\ref{b90}}$ of inflationary
theory can be realized.
Several years ago it was
realized that inflation in these theories has a very interesting
property$^{\ref{b19},\ref{b20}}$ which  will be discussed in this talk.
If the Universe contains at least one inflationary domain of a size
of horizon
($h$-region) with a sufficiently large and homogeneous scalar field
$\phi$, then
this domain will permanently produce new $h$-regions of a similar
type. During
this process the total physical volume of the inflationary Universe
(which is
proportional to the total number of $h$-regions) will grow
indefinitely.

 Fortunately, some
kind of stationarity may exist in many models of  inflationary
Universe due to
the process of the Universe self-reproduction$^{\ref{MyBook}}$.  The
properties of
inflationary domains formed during the process of the
self-reproduction of the
Universe do not depend on the moment of time at which each such
domain is
formed; they depend only on the value of the scalar fields inside
each domain, on
the average density of matter  in this domain and on the physical
length scale. This kind
of stationarity, as opposed to the ``stationary distribution''
$P_c(\phi,t)$ (the probability density to find the field $\phi$ inside a
given Hubble domain at time $t$) cannot be described in the
minisuperspace approach.

In order to describe the structure of the inflationary Universe beyond one
$h$-region (minisuperspace) approach one has to investigate the probability
distribution $P_p(\phi,t)$, which  takes into account the inhomogeneous
exponential growth of the volume of domains filled by field
$\phi$$^{\ref{b19}}$.  Solutions for this probability distribution were
first
obtained in$^{\ref{b19},\ref{b20}}$ for the case of chaotic inflation. It
was shown  that if the initial value of the scalar field $\phi$ is
greater than
some critical value $\phi^*$, then the probability distribution
$P_p(\phi,t)$
permanently moves to larger and larger fields $\phi$, until it
reaches the field $\phi_p$, at which the effective potential of the field
becomes of the order of Planck density $M_p^4$ (we will assume $M_p=1$
hereafter), where the standard methods of quantum field theory in a curved
classical space are no longer valid. By the methods used
in$^{\ref{b19},\ref{b20}}$
it was impossible to check whether $P_p(\phi,t)$ asymptotically approaches
any
stationary regime in the classical domain $\phi < \phi_p$.

Several important steps towards the solution of this problem were
made  by Aryal and Vilenkin$^{\ref{ArVil}}$,  Nambu and
Sasaki$^{\ref{Nambu}}$
and Miji\'c$^{\ref{Mijic}}$. Their papers contain many beautiful results
and
insights, and we will use many results obtained by these authors. However,
the generalization of the results of$^{\ref{ArVil}}$ to non-trivial case of
chaotic inflation appeared to be not an easy
problem$^{\ref{LLMPascos},\ref{MezhLin}}$\@.  Miji\'c$^{\ref{Mijic}}$  did
not have a
purpose to obtain a complete expression for the stationary distribution
$P_p(\phi,t)$\@. The corresponding expressions were obtained for various
types
of potentials $V(\phi)$ in$^{\ref{Nambu}}$. Unfortunately, according
to$^{\ref{Nambu}}$, the stationary distribution $P_p(\phi,t)$ is almost
entirely concentrated at $\phi \gg \phi_p$, i.e. at $V(\phi) \gg 1$, where
the
methods used in$^{\ref{Nambu}}$ are inapplicable.

In this talk we will argue that the process of the
self-reproduction of inflationary Universe
effectively kills itself at densities approaching
the Planck
density. This leads to the existence of a
 stationary probability distribution $P_p(\phi,t)$ concentrated
entirely at
sub-Planckian densities $V(\phi) < 1$  in a wide class of
theories leading
to chaotic inflation.

Let us consider the simplest
model of chaotic inflation based
on the  theory of a  scalar field $\phi$ minimally coupled to
gravity, with the
Lagrangian
\begin{equation}\label{E01}
L =  \frac{1}{16\pi}R + \frac{1}{2} \partial_{\mu} \phi
\partial^{\mu} \phi - V(\phi)  \ .
\end{equation}
Here $G = M^{-2}_p = 1$ is the gravitational constant,  $R$ is the
curvature
scalar, $V(\phi)$ is the effective potential of the scalar field. If
the classical
field $\phi$ is sufficiently homogeneous in some domain of the
Universe (see
below), then its behavior inside this domain is governed by the
equations
\begin{equation}\label{E02}
\ddot\phi + 3H\dot\phi = -dV/d\phi,
\end{equation}
\begin{equation}\label{E03}
H^2 + \frac{k}{a^2} = \frac{8\pi}{3}\, \left(\frac{1}{2}
\dot\phi^2 + V(\phi)\right) \ .
\end{equation}
Here $H={\dot a}/a, a(t)$ is the scale factor of the
 Universe, $k=+1, -1,$ or $0$ for a closed, open or
flat Universe, respectively.

The most important fact for inflationary scenario is that for most
potentials $V(\phi)$ (e.g., in all power-law $V(\phi)=g_n \phi^n /n$ and
exponential $V(\phi)=g e^{\alpha \phi}$ potentials) there is an
intermediate
asymptotic regime of slow rolling of the field $\phi$ and quasi-exponential
expansion of the Universe. This expansion (inflation)
ends at $\phi \sim \phi_e$ where the slow-rolling regime $\ddot\phi \ll
3H(\phi) \dot\phi$ breakes down.

If, as it is usually assumed, the classical description of the
Universe becomes possible only when the energy-momentum
tensor of matter becomes smaller than $1$, then at this
moment $\partial_{\mu}\phi\partial^{\mu}\phi \la 1$ and $V(\phi) \la
1$\@. Therefore, the only constraint on the initial amplitude of
the field $\phi$ is given by $V(\phi) \la 1$.
This gives a typical initial value of the field $\phi$ in the theory
(\ref{E01}):
\begin{equation}\label{E08'}
\phi_0 \sim  \phi_p \ ,
\end{equation}
where $\phi_p$ corresponds to the Planck energy density, $V(\phi_p) =
1$.

During the inflation all the inhomogeneities are stretched away and, if the
evolution of the Universe were governed solely by classical equations of
motion (\ref{E02}), (\ref{E03}), we would end up with extremely smooth
geometry of the spatial section of the Universe with no primordial
fluctuations to initiate the growth of galaxies and large-scale structure.
Fortunately, the same Jeans instability which causes the growth of galaxies
during the Hot Big Bang era leads to the existence of the growing modes of
vacuum fluctuations during the inflation. The wavelengths of all  vacuum
fluctuations of the scalar field $\phi$ grow exponentially
in the expanding Universe. When the wavelength of any
particular fluctuation becomes greater than $H^{-1}$, this
fluctuation stops oscillating, and its amplitude freezes at
some nonzero value $\delta\phi (x)$ because of the large
friction term $3H\dot{\phi}$ in the equation of motion of the field
$\phi$\@. The amplitude of this fluctuation then remains
almost unchanged for a very long time, whereas its
wavelength grows exponentially. Therefore, the appearance of
such a frozen fluctuation is equivalent to the appearance of
a classical field $\delta\phi (x)$ that does not vanish
after averaging over macroscopic intervals of space and
time.

Because the vacuum contains fluctuations of all
wavelengths, inflation leads to the creation of more and
more perturbations of the classical field with
wavelengths greater than $H^{-1}$\@. The average amplitude of
such perturbations generated during a time interval $H^{-1}$
(in which the Universe expands by a factor of e) is given
by
\begin{equation}\label{E23}
|\delta\phi(x)| \approx \frac{H}{2\pi}\ .
\end{equation}
 If the field is massless (or, better to say, as long as $m^2(\phi) \ll
H^2(\phi)$), the amplitude of each frozen wave does not change in time at
all.
On the other hand, phases of each waves are random.
Therefore,  the sum of all waves at a given point fluctuates and
experiences
Brownian jumps in all directions in the space of fields.

The standard way of description of the stochastic behavior of the inflaton
field during the slow-rolling stage is to coarse-grain it over $h$-regions
and
consider the effective equation of motion of the long-wavelength
field$^{\ref{b60},\ref{b61},\ref{Star}}$:
 \begin{equation} \label{m1}
\frac{d}{dt} \, \phi = - \frac{V'(\phi)}{3H(\phi)} +
\frac{H^{3/2}(\phi)}{2\pi}
\, \xi(t) \ ,
\end{equation}
Here $\xi(t)$ is the effective white noise generated by quantum
fluctuations,
which leads to the Brownian motion of the classical field $\phi$\@.

This Langevin equation corresponds to the following Fokker-Planck equation
for the probability density $P_c(\phi,t)$ to find the field $\phi$ in a
given point (which now means $h$-region) after time $t$:
\begin{equation}\label{E3711}
\frac{\partial P_c}{\partial t} =
\frac{\partial}{\partial \phi} \left({H^{3/2}(\phi)\over
8\pi^2} \
\frac{\partial}{\partial \phi}\, \left(H^{3/2}(\phi) P_c \right) +
{V'(\phi)\over 3H(\phi)} \, P_c\right) \ ,
\end{equation}

The formal stationary solution ($\partial P_c/\partial t=0$) of
equation (\ref{E3711}) would be$^{\ref{b60}}$
\begin{equation}\label{E38}
P_c \sim \exp\left({3\over 8 V(\phi)}\right) \ ,
\end{equation}

Note, that the stationary solution (\ref{E38}) is equal to the square
of the
Hartle-Hawking wave function of the Universe$^{\ref{HH}}$.  At  first
glance, this
result is a direct confirmation of the Hartle-Hawking
prescription for the wave function of the Universe.

However, in all realistic cosmological theories, in which $V(\phi)=0$
at its
minimum, the distribution (\ref{E38}) is not
normalizable. The source of this difficulty can be easily
understood: any stationary distribution may exist only due
to a compensation of a classical flow of the field $\phi$
downwards to the minimum of $V(\phi)$ by the diffusion motion
upwards. However, diffusion of the field $\phi$ discussed
above exists only during inflation, i.e. only for $\phi \geq
1$, $V(\phi)\geq V(1)\sim m^2\sim 10^{-12}$ for $m \sim 10^{-6}$.
Therefore (\ref{E38}) would correctly describe the stationary
distribution $P_c(\phi)$ in the inflationary Universe only if
$V(\phi)\geq 10^{-12} \sim 10^{80}$ GeV in the absolute minimum of
$V(\phi)$, which is, of course, absolutely unrealistic$^{\ref{b20}}$.

It was shown$^{\ref{b20}}$ that the solutions of this equation with the
effect
of ``end of inflation'' boundary properly taken into account are
non-stationary (decaying). It is of no surprize because we didn't take into
account yet the complicated and inhomogeneous expansion of the whole
Universe with multiplicating number of $h$-regions. The close view on the
process of the growth of the volume of inflationary Universe reveals the
strong resemblance with branching diffusion processes, with the role of
branching particles being played by $h$-regions and the the diffusing
parameter being associated with the inflaton field inside an $h$-region. If
one wishes to describe the distribution of the inflaton field in the whole
Universe and not only in one Hubble domain, one has to modify the
Fokker-Planck equation by introducing the term corresponding to creation of
more and more new ``particles'' during the diffusion process
\ref{LLMPascos},\ref{MezhLin},\ref{Nambu},\ref{ZelLin}\@.
\begin{equation}\label{E372}
\frac{\partial P_p}{\partial t} = \frac{\partial }{\partial\phi}
 \left( {H^{3/2}(\phi)\over 8\pi^2} \frac{\partial }{\partial\phi}
\left(
 {H^{3/2}(\phi)}P_p \right)
 +  \frac{V'(\phi)}{3H(\phi)} \, P_p \right)
 +  3H(\phi)  P_p
\end{equation}

The mathematical model describing such behavior is
the recently developed theory of a new type of branching diffusion
processes$^{\ref{MezhMolch}}$\@. Eq.\ (\ref{E372}) is, from this
perspective,
the forward Kolmogorov equation for the first moment of the generating
functional of number of branching particles.

There are two main sets of questions which may be asked concerning such
processes. First of all, one may
be interested in the probability $P_p(\phi, t)$ to find a given field
$\phi$ at a
given time $t$ under the condition that initial value of the field
was equal to
some $\phi(t=0) = \phi_0$\@. In what follows we will denote $\phi_0$
as $\chi$\@. On the other hand, one may wish to know, what is the
probability
$P_p(\chi,t)$  that the given final value of the field $\phi$ (or the state
with a given final density $\rho$) appeared as a process of diffusion and
branching of a domain containing some field $\chi$\@. Or, more generally,
what
are the typical history of branching Brownian trajectories which end up at
a
hypersurface of a given $\phi$ (or a given $\rho$)?

It happens that if we do not take the Planck boundary seriously, this
equation also doesn't have a true stationary solution (the solutions found
by Nambu {\it et.\ al.\ }$^{\ref{Nambu}}$ are heavily concentrated at the
super-Planckian densities which is just another way to state that there is
no
real stationarity). This conclusion, however, may fail if we will treat the
Planck boundary more carefully. There are different reasons to do this:

{\bf 1.} Eq. (\ref{E372}) may have a slightly different form, which
corresponds to the difference between stochastic approaches of Ito
and
Stratonovich. This difference is not important at densities much
smaller than $1$, but at $V(\phi) \ga 1$ it may become significant.
(Note,
however, that this is not an unsolvable problem. One may just
investigate modified equations as well. Two other problems are more
fundamental.)

{\bf 2.} Diffusion equations were derived in the semiclassical
approximation
which breaks down near the Planck energy density $\rho_p \sim M^4_p =
1$.

{\bf 3.} Interpretation of the processes described by these equations is
based on the notion of classical fields in a classical space-time,
which is not applicable at densities larger than $1$ because of large
fluctuations of metric at such densities. In particular, our
interpretation of $P_c$ and $P_p$ as of probabilities to find
classical
field $\phi$ in a given point at a given time does not make much
sense
at $\rho > 1$.

There is also another, more general, reason to expect the existence of the
stationary solutions: as we will argue now, inflation kills itself as the
density approaches the Planck density $\rho_p \sim M_p^4$.

In our previous investigation we assumed that the vacuum energy
density is
given by $V(\phi)$ and the energy-momentum tensor is given by
$V(\phi)\, g_{\mu\nu}$\@. However, quantum fluctuations of the scalar
field give the contribution to the average value of the energy
momentum
tensor, which does not depend on mass (for $m^2 \ll H^2$) and is
given
by$^{\ref{b99}}$
\begin{equation}\label{x1}
 <T_{\mu\nu}> \, = \, {3\, H^4\over 32\, \pi^2}\ g_{\mu\nu} \, =\,
 {2 \over 3}\, V^2 \ g_{\mu\nu} \ .
\end{equation}
The origin of this contribution is obvious. Quantum
fluctuations of the scalar field $\phi$ freeze out with the amplitude
${H\over 2\pi}$ and the wavelength $\sim H^{-1}$\@. Thus, they lead to
the gradient energy density $(\partial_\mu\delta\phi)^2\sim H^4$.

An interesting property of eq. (\ref{x1}) is that the average value
of the
energy-momentum tensor of quantum fluctuations does not look  as an
energy-momen\-tum tensor corresponding to the gradients  of a
sinusoidal wave.
It looks rather as a renormalization of the vacuum energy-momentum
tensor (it
is proportional to $g_{\mu\nu}$)\@. This means, in particular, that
after averaging over all possible outcomes of the process of generation of
long-wave
perturbations, the result (for $<T_{\mu\nu}>$) does not depend on the
choice of
the coordinate system. However, if we are interested in local events (i.e.\
we are averaging only over short wavelengths),
 we will see long-wave inhomogeneities produced by the ``frozen''
fluctuations of  the scalar field. This does not lead to any interesting
effects at $V \ll 1$ ($\phi \ll \phi_p$)  since in this case $V^2 \ll V$.

However, at the density larger than the Planck density  the situation
becomes much more complicated. At $V > 1$ the gradient energy density
$\sim
V^2$ becomes  larger than the potential energy density $V(\phi)$\@. A
typical
wavelength of perturbations giving the main contribution to the
gradient
energy is given by the size of the horizon, $l \sim H^{-1}$\@. This
means that the
inflationary Universe at the Planck density becomes divided into many
domains
of the size of the horizon, density contrast between each of these
domains
being of the order of one. These domains evolve as separate
mini-Universes with
the energy density dominated not by the potential energy density but
by the
 energy density of gradients of the field $\phi$\@. Such domains
drop out from the process of exponential expansion. Some of them may
re-enter
this process later, but many of them will collapse within the typical
time $H^{-1}$.

This can be effectively described by imposing some kind of absorbing (or
reflecting, or elastic screen type) boundary conditions on our process of
branching diffusion at some $\phi_b$ which should be close to the Planck
boundary. Without going into details of derivation, we just announce here
the
results$^{\ref{MezhLin}}$.

Our calculations have shown$^{\ref{MezhLin}}$ that if our
boundary conditions do not permit the field $\phi$
penetrate deeply into the domain $\phi > \phi_p$,
the final results of our investigation do not depend on the
type of the boundary conditions imposed (whether they are absorbing,
reflecting, etc.)\@. They depend only on the value of the field $\phi_b$
where the boundary condition is to be imposed, and this dependence is
rather trivial. We have argued that $\phi_b \sim \phi_p$\@.
Therefore we will assume now that the function $P_p(\phi,t|\chi)$
satisfies boundary conditions corresponding to disappearing particles at
$\phi_p$ treating $\phi_p$ now just as a phenomenological parameter not
necessarily corresponding to $V(\phi_p)=1$\@.   In this case eq.\
(\ref{E372}) and its conjugate (backward Kolmogorov equation) are endowed
by
the following condition:
\begin{equation} \label{eq8a}
P_p(\phi_p,t|\chi) =  0\ .
\end{equation}

One may try to obtain solutions of equations (\ref{E372}) and its conjugate
in
a form of the following series of biorthonormal system of eigenfunctions
of the pair of adjoint linear operators (defined by the equations below):
\begin{equation} \label{eq14}
P_p(\phi_p,t|\chi) =
\sum_{s=1}^{\infty} { e^{\lambda_s t}\, \Pes{s}\, \Dpi{s} } \ .
\end{equation}
Indeed, this gives us a solution of eq. (\ref{E372}) if
\begin{equation} \label{eq15}
\frac{1}{2} \frac{H^{3/2}(\chi)}{2\pi} \frac{\partial}{\partial\chi}
 \left( \frac{H^{3/2}(\chi)}{2\pi} \frac{\partial }{\partial\chi}
\Pes{s} \right) - \frac{V'(\chi)}{3H(\chi)} \frac{\partial }{\partial\chi}
\Pes{s}
+ 3H(\chi) \, \Pes{s} = \lambda_s \, \Pes{s}  \ .
\end{equation}
\begin{equation} \label{eq17}
\frac{1}{2}  \frac{\partial }{\partial\phi}
 \left( \frac{H^{3/2}(\phi)}{2\pi} \frac{\partial }{\partial\phi}
\left(
 \frac{H^{3/2}(\phi)}{2\pi} \Dpi{j} \right) \right)
 + \frac{\partial }{\partial\phi} \left( \frac{V'(\phi)}{3H(\phi)} \,
\Dpi{j} \right)
+ 3H(\phi)  \, \Dpi{j} =
\lambda_j \, \Dpi{j}  \ .
\end{equation}
The orthonormality condition reads
\begin{equation} \label{eq20}
\int_{\phi_e}^{\phi_p} { \Pes{s} \, \pi_j(\chi) \, d\chi }
= \delta_{sj}
\end{equation}

In our case (with regular boundary conditions) one can easily show
that the
spectrum of $\lambda_j$ is discrete and bounded from above. Therefore
the
asymptotic solution for $P_p(\phi,t|\chi)$ (in the limit $t
\rightarrow \infty$) is given by
\begin{equation} \label{eq22}
P_p(\phi_p,t|\chi) = e^{\lambda_1 t}\, \Pes{1} \,
\Dpi{1}\, \cdot \left(1 + O\left( e^{-\left(\lambda_1 - \lambda_2
\right) t} \right) \right) \ .
\end{equation}
Here $\Pes{1}$ is the only positive eigenfunction of eq.
(\ref{eq15}),
$\lambda_1$ is the corresponding (real) eigenvalue, and $\pi_1(\phi)$
(invariant density of branching diffusion) is the eigenfunction
of the conjugate operator (\ref{eq17}) with the
same eigenvalue $\lambda_1$\@. Note, that $\lambda_1$ is
the largest eigenvalue, $\mbox{Re} \left( \lambda_1 - \lambda_2
\right) > 0 $\@.  We found$^{\ref{MezhLin}}$ that
in realistic theories of inflation the typical time of
relaxation to the asymptotic
regime, $t_{rel} \sim (\lambda_1 - \lambda_2)^{-1}$, is extremely
small. It is only about a few thousands Planck times, i.e. about
$10^{-40}~sec$\@. This means, that the normalized distribution
\begin{equation} \label{eq22aa}
\tilde{P}_p(\phi,t|\chi) = e^{-\lambda_1 t}
\,P_p(\phi_p,t|\chi)
\end{equation}
rapidly converges to the time-independent normalized distribution
\begin{equation} \label{eq22a}
\tilde{P}_p(\phi|\chi) \equiv
\tilde{P}_p(\phi,t \rightarrow \infty|\chi) =  \Pes{1} \, \Dpi{1} \ .
\end{equation}
It is this stationary distribution that we were looking for. The
remaining problem is to find the functions $\Pes{1}$ and $\Dpi{1}$,
and to check that all assumptions about the boundary conditions which
we made on the way to eq. (\ref{eq22}) are actually satisfied.

After some calculations we came to the following expression for
$P_p(\phi|\chi)$ (note that the functions $\Phi(\cdot)$ are essentially the
same, the only difference is the argument):

\begin{equation} \label{eq32}
P_p(\phi|\chi) = \frac{C}{V^{3/4}(\phi)} \, \exp \left\{
\frac{3}{16\, V(\phi)}
- \frac{3}{16\, V(\chi)} \right\} \, \Phi(z(\chi)) \,
\Phi(z(\phi))
\end{equation}
Here the normalization constant $C$ should be determined from
(\ref{eq20})\@. Using the WKB approximation, we have
calculated$^{\ref{MezhLin}}$ $\Phi(z(\phi))$  for a wide class of
potentials
usually considered in the context of chaotic inflation, including
potentials
$V \sim  \phi^n$ and $V \sim e^{\alpha\phi}$\@.

The solution we have found features interesting properties. First
of all, it is concentrated heavily at the highest allowed values
of the inflaton field. Despite the promising exponential prefactor
in (\ref{eq32}) which looks like a combination of the
Hartle-Hawking$^{\ref{HH}}$ and tunneling$^{\ref{tunnel}}$
wavefunctions, the dependence of $\Phi(z(\phi))$ on $\phi$ appeared to be
even
more steep that those exponents. This function falls exponentially (with
the
rate governed by the Planckian energies) towards the lower values of the
inflaton field and strongly overwhelms the dependence of the exponent in
front
of it. Only near the ``end of inflation'' boundary the solution
(\ref{eq32})
reveals something familiar --- the dependence on the initial field $\chi$
becomes similar to the square of tunneling wavefunction  (simultaneously,
the
corresponding dependence on the final field $\phi$ cancels out)\@. And, as
we
have already mentioned, the relaxation time is very small (which is also a
consequence of the fact that the dynamics of the self-reproducing Universe
is
governed by the maximal possible energies)\@. The
stationary distribution found in$^{\ref{MezhLin}}$ is  not very
sensitive to our  assumptions concerning the concrete mechanism of
suppression  of production  of inflationary domains with $\phi \ga
\phi_p$\@.   We hope that these results may show us a way towards the
complete
quantum mechanical description of
the stationary ground state of the Universe.

A.M. is grateful to A.Linde and D.Linde for fruitful and enjoyable
collaboration on this project and to A.Starobinsky, S.Molchanov and
A.Vilenkin for discussions.

\vskip 1.5cm
\centering{\Large \bf References}
\begin{enumerate}
\item \label{LLMPascos} A.Linde, D.Linde, A.Mezhlumian, {\it From the Big
Bang
Theory to the Theory of a Stationary Universe}, Stanford preprint to
appear.
\item \label{MezhLin} A. Linde and A. Mezhlumian, Stanford preprint to
appear.
\item \label{b17} A.D. Linde, Phys. Lett. {\bf 129B} (1983) 177.
\item \label{b16} A.D. Linde, Phys. Lett. {\bf 108B} (1982); {\bf 114B}
(1982) 431; {\bf 116B} (1982) 335, 340;  A. Albrecht and P.J. Steinhardt,
Phys. Rev. Lett. {\bf 48} (1982) 1220.
\item \label{b90} D. La and P.J. Steinhardt, Phys. Rev. Lett. {\bf 62}
(1989) 376.
\item \label{b19} A.D. Linde, Phys. Lett. {\bf 175B} (1986) 395; Physica
Scripta
{\bf T15} (1987) 169. 
\item \label{b20} 	A.S. Goncharov and A.D. Linde, Sov. Phys. JETP {\bf
65} (1987) 635; A.S. Goncharov, A.D. Linde and V.F. Mukhanov, Int. J. Mod.
Phys. {\bf A2} (1987) 561.
\item \label{MyBook} A.D. Linde, {\bf Particle Physics and Inflationary
Cosmology} (Harwood, Chur, Switzerland, 1990);
A.D. Linde, {\bf Inflation and Quantum Cosmology}
(Academic Press, Boston, 1990).
\item \label{ArVil} M. Aryal and A. Vilenkin, Phys. Lett. {\bf B199}
(1987) 351.
\item \label{Nambu} Y.Nambu and M.Sasaki, Phys.Lett. {\bf B219} (1989) 240;

Y. Nambu, Prog. Theor. Phys. {\bf 81} (1989) 1037.
\item \label{Mijic} M. Miji\' c, Phys. Rev. {\bf D42} (1990) 2469; Int.
J. Mod. Phys. {\bf A6} (1991)  2685.
\item \label{b60} A.A. Starobinsky, in: {\bf Fundamental Interactions}
(MGPI Press, Moscow, 1984), p. 55.
\item \label{b61} A.S. Goncharov and A.D. Linde, Sov. J. Part. Nucl.
  {\bf 17} (1986) 369.
\item \label{Star} A.A. Starobinsky, in: {\bf Current Topics in Field
Theory, Quantum Gravity and Strings}, Lecture Notes in Physics, eds.
H.J. de Vega and N. Sanchez (Springer, Heidelberg 1986) {\bf 206},
p. 107.
\item \label{b99} T.S. Bunch and P.C.W. Davies, Proc. Roy. Soc. {\bf
A360}
(1978) 117; A. Vilenkin and L. Ford, Phys. Rev. {\bf D26} (1982)
1231;
A.D. Linde, Phys. Lett. {\bf 116B} (1982) 335;
A.A. Starobinsky, Phys. Lett. {\bf 117B} (1982) 175;
A. Vilenkin, Nucl. Phys. {\bf B226} (1983) 527.
\item \label{HH} J.B. Hartle and S.W. Hawking, Phys. Rev. {\bf D28}
(1983) 2960.
\item \label{ZelLin} Ya.B. Zeldovich and A.D. Linde, unpublished (1986).
\item \label{MezhMolch} A. Mezhlumian and S.A. Molchanov,  Stanford
University preprint SU-ITP-92-32 (1992), available as cond-mat/9211003.
\item \label{tunnel} A.D. Linde, JETP {\bf 60} (1984) 211; Lett. Nuovo Cim.
{\bf 39} (1984) 401; Ya.B. Zeldovich and A.A. Starobinsky, Sov. Astron.
Lett.{\bf 10} (1984) 135; V.A. Rubakov, Phys. Lett. {\bf 148B} (1984) 280;
A.
Vilenkin, Phys. Rev. {\bf D30} (1984) 549.

 \end{enumerate}

\end{document}